\documentclass[prl,aps,twocolumn,showpacs]{revtex4}

\usepackage[dvips]{graphicx}
\usepackage{dcolumn}%
\usepackage{amsmath}%
\usepackage{amsfonts}%
\usepackage{amssymb}
 \usepackage[usenames]{color}
\usepackage{appendix}

\providecommand{\U}[1]{\protect\rule{.1in}{.1in}}
\newcommand{\be}{\begin{equation}}
\newcommand{\ee}{\end{equation}}
\newcommand{\bea}{\begin{eqnarray}}
\newcommand{\eea}{\end{eqnarray}}
\newcommand{\bt} {\begin{tabular}}
\newcommand{\et} {\end{tabular}}
\newcommand{\nn}{ \nonumber}
\newcommand{\ds}{\displaystyle}
\newcommand{\ba} {\begin{array}}
\newcommand{\ea} {\end{array}}
\begin{document}

\title{Fano effect in a thermally induced transport through a triple quantum dot within the Coulomb blockade regime}

\author{  Natalya A. Zimbovskaya{\footnote{Corresponding author: natalya.zimbovskaya@upr.edu}}}

\affiliation
{Department of Physics and Electronics, University of Puerto Rico-Humacao, CUH Station, Humacao, PR 00791, USA}

\begin{abstract}
 In the present work we theoretically study thermoelectric transport and heat transfer in a junction including a $T$-shaped double quantum dot coupled to nonmagnetic electrodes and supplemented  with a third dot in a parallel configuration. We focus on the combined effect of Coulomb interactions and quantum interference occurring in the $T$-shaped portion of the considered triple quantum dot on the thermoelectric electron transport. The transport through the system is studied within the Coulomb blockade regime in the limit of strong intra-dot interactions between electrons beyond the linear response regime. It is shown that under these conditions both charge and heat transfer through the considered system may be significantly affected by the quantum interference and inter-dot Coulomb interactions. 
\end{abstract}

\date{\today}
\maketitle

\section{I. Introduction.}
 
 In the last two decades, transport properties of tailored nanoscale systems including single and/or multiple quantum dots (QD) are being intensively studied due to their fundamental and applied perspectives \cite{1,2,3,4,5}. Advances in controlling and measurements of heat transferred through small systems \cite{6,7,8} further intensified these studies which suggest new possibilities  for manufacturing nanoscale devices such as heat-to-electric energy converters \cite{9,10,11,12}, cooling systems \cite{2,3,13,14} and heat diods \cite{15,16,17,18,19,20,21,22,23}.

 One of the key features of transport through a QD is a high degree of phase coherence being preserved during electron transfer. Provided that several transport channels for electrons coexist, interference between different paths can be manifested through various interference effects such as Aharonov-Bohm oscillations in the electron conductance \cite{24,25,26,27}, Fano resonances/antiresonances \cite{28,29,30,31} and Dicke effect \cite{32}. The pattern may be affected by the interplay between the charge/heat and spin transport through the system with ferromagnetic \cite{33,34,35,36} or superconducting \cite{36,37,38,39} electrodes. Also, Coulomb interactions between electrons on the dot as well as inter-dot Coulomb interactions in multiple QDs may affect the interference pattern in both Coulomb blockade regime \cite{32,40,41,42} and Kondo regime \cite{43,44,45,46}.

In the present work we theoretically analyze the influence of Fano effect and Coulomb interactions on thermoelectric transport through multiple QDs beyond the linear response regime. We choose a simple model representing a triple quantum dot placed in between normal nonmagnetic electrodes shown in Fig.1.  The triple QD includes dot 1 linking the left and right electrodes, side dot 2 detached from the electrodes but coupled to dot 1 (these dots form a T-shaped double dot providing two different paths for traveling electrons which may lead to emergence of quantum interference effects) and dot 3 attached to the right electrode.  The chosen model resembles those used in some earlier works \cite{31,42} where multiple QD including side dots were used to study manifestations of Fano effect in thermoelectric transport characteristics within the linear response regime. Here, we supplement the T-shaped double dot with dot 3 which creates a spatial asymmetry in the system thus providing opportunities for rectification of charge and heat currents to occur.

 The analysis is carried on using nonequilibrium Green's functions formalism. It is simplified by the assumption that phonon contribution to the electron transport through QDs is small, as often happens in real systems due to the mismatch between frequencies of phonon modes associated with the dots embedded in their matrices and modes associated with the environment preventing the overlap between these groups of phonons \cite{47,48}. Basing on this assumption, we omit from consideration phonon contribution to the heat transfer and all effects originating from electron-phonon interactions. We assume that there exist both intra-dot and inter-dot Coulomb interactions between electrons concentrating on the limit of strong intra-dot electron-electron interactions. We show that within this limit inter-dot Coulomb interactions together with Fano resonances occurring in the T-shaped portion of the triple dot may bring significant changes into charge and heat currents.
\begin{figure}[t] 
\begin{center}
\includegraphics[width=8cm,height=6cm]{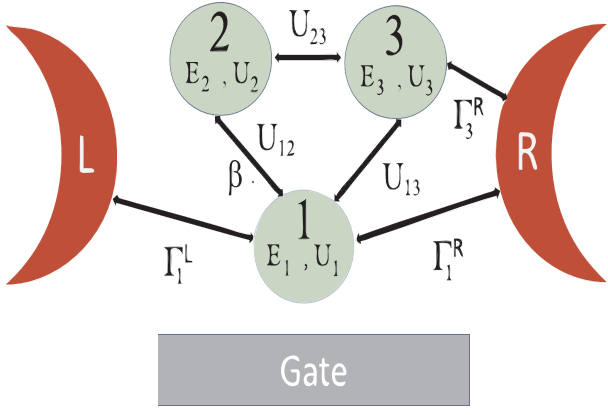} 
\caption{Schematics of the considered transport junction including a T-shaped double dot placed between electrodes and comprised of the dot 1 and dot 2 and a third dot (dot 3) attached to the right electrode.
}
 \label{rateI}
\end{center}\end{figure}

The paper is organized as follows. In Sec.II we describe the employed model and the formalism. In Sec.III and Sec.IV we apply obtained results to analyze the effect of Fano interference on the charge and heat currents, respectively. Conclusions are given in Sec.V. 

\section{II. Model}

As mentioned above, we omit electron-phonon interactions and assume that the coherent electron transmission is the predominant mechanism  governing electron transport. Within Anderson model, the considered system is described by the Hamiltonian:
\be
H=H_E+H_d+H_t        \label{1}
\ee
Here, the contribution from the electrodes has the form:
\be
H_E=\sum_{\alpha,k,\sigma}\epsilon_{\alpha,{\bf k},\sigma}C^{\dag}_{\alpha,{\bf k},\sigma}C_{\alpha,{\bf k},\sigma},        \label{2}
\ee
where$\alpha=\{L,R\}$ labels the left and right electrodes, $\epsilon_{\alpha,{\bf k},\sigma}$ are single electron energies in the electrodes and $C^{\dag}_{\alpha,{\bf k},\sigma}, C_{\alpha,{\bf k},\sigma}$ are creation and annihilation operators for electron states ${\bf k}, \sigma$ (the index $\sigma$ is labeling spin-up and spin-down electrons).

The term $H_d$  is associated with the triple dot:
\begin{align}
H_d=&\sum_{i,\sigma}n_{i,\sigma}(E_{i}+U_{i}n_{i,-\sigma})+\sum_{\sigma,\sigma'}\sum_{i,j}U_{ij}n_{i,\sigma}n_{j,\sigma'}
\nn\\&
+\beta\sum_{\sigma}\left(d^{\dagger}_{1,\sigma}d_{2,\sigma}+H.C\right).       \label{3}
\end{align}
In this expression, $1\leq i,j\leq 3$ $(i\neq j$), $E_{i}$ are electron energies on the dots, $ U_{i}$ and $U_{ij}$ are intra-dot and inter-dot charging energies, $n_{i,\sigma}=d^{\dag}_{i,\sigma}d_{i,\sigma}$, and $d^{\dag}_{i,\sigma}, d_{i,\sigma}$ are creation and annihilation operators for electrons on the dots. Within the chosen model electron tunneling characterized by the coupling parameter $\beta$ occurs solely between dot 1 and  side dot 2. In the following calculations, Coulomb energies $U_{ij}$ corresponding to interactions between dot 3 and dots 1 and 2 are presumed to be equal ($U_{i,3}=U_{3,i}=U$) for $i=\{1,2\}$ as well as all intra-dot Coulomb energies ($U_i=\tilde U$). Also, we put $U_{12}=U_{21}=U_0$.

The last term which describes the couplings of the triple dot to the electrodes has the form:
\be
H_t=\sum_{{\bf k},\sigma,i} \left(t_{L,{\bf k},\sigma,i}C^{\dag}_{L,{\bf k},\sigma}d_{i,\sigma}+t_{R,{\bf k},\sigma,i}C^{\dag}_{R,{\bf k},\sigma}d_{i,\sigma}\right)+H.C.        \label{4}
\ee
 where $i=1,3$ and factors $t_{\alpha,{\bf k},\sigma,i}$ represent coupling strengths between the dots and the electrodes. Note that there is no electron transfer between electrodes and the side dot 2.

Basing on the Hamiltonian (\ref{1}) we compute expressions for the retarded (${\bf G}^{r}_{\sigma}(E)$) and advanced (${\bf G}^{a}_{\sigma}(E)$) electron Green's functions for the considered triple dot. We use the version of equations of motion (EOM) method suggested in Refs.\cite{49,50}. The Green's functions are computed within the Coulomb blockade regime where charging energies significantly exceed thermal energies $kT_{\alpha}$ ($T_{\alpha}$ being the electrodes temperatures and $k$ being the Boltzmann constant) as well as energies characterizing couplings of the dots to electrodes. We consider the limit of strong intra-dot Coulomb interactions assuming that $\tilde U\gg U,U_{0},E_{i},\beta$ and  moderately biased systems $|\Delta\mu=\mu_L-\mu_R|<\tilde U$ ($\mu_{L,R}$ being  chemical potentials of the electrodes). Under such conditions each dot may contain only a single electron because energy levels $E=E_{i}+\tilde U$ are shifted beyond the conduction window and remain empty because electrons tunnel from there to the electrodes. As a result, strong intra-dot Coulomb interactions become nearly irrelevant whereas much weaker inter-dot Coulomb interactions may significantly affect electron transport. 

Under the accepted conditions expressions for nonzero matrix elements $ G^{r}_{ij\sigma}(E)$ may be reduced to the form ($\{i,j\}=\{1,2\}$, $i\neq j$):
\be
G^{r}_{ii,\sigma}(E)=\frac{\upsilon_{i,\sigma}}{1-\beta^2\upsilon_{i,\sigma}(E)\upsilon_{j,\sigma}(E)};  \label{5}
\ee
\be
G^{r}_{ij,\sigma}(E)=\frac{\beta\upsilon_{i,\sigma}\upsilon_{j,\sigma}}{1-\beta^2\upsilon_{i,\sigma}(E)\upsilon_{j,\sigma}(E)};  \label{6}
\ee
and
\be
G^{r}_{33,\sigma}(E)=\upsilon_{3,\sigma}               \label{7}
\ee.
In these expressions \cite{15}: 
\be
\upsilon_{i,\sigma}(E)= \left(1-\big<n_{i,-\sigma}\big>\right)\sum_{m=1}^{4}\frac{P_m}{E-E_i-\Pi_{m}+i\Gamma_i}    \label{8}
\ee
where $1\leq i,j,k\leq 3$, $j\neq\ k\neq i$,
$P_1=(1-\big<n_j\big>)(1-\big<n_k\big>)$, $P_2=\big<n_j\big>(1-\big<n_k\big>)$, $P_3=\big<n_k\big>(1-\big<n_j\big>)$, $P_4=\big<n_j\big>\big<n_k\big>$, $\big<n_j\big>=\big<n_{j,\sigma}\big>+\big<n_{j,-\sigma}\big>$, $\Pi_1=0$, $\Pi_2=U_0$, $\Pi_3=U$ and $\Pi_4= U_{0}+U$. Electron occupation numbers on the dots $\big<n_{i,\sigma}\big>$ are determined by the equations:
\be
\big<n_{i,\sigma}\big>=\frac{1}{2\pi i}\int dE G^{<}_{ii,\sigma}(E).                                       \label{9}
\ee

 The lesser Green's function ${\bf G}^{<}_{\sigma}(E)$ may be approximated as ${\bf G}^{<}_{\sigma}(E)=i\sum_{\alpha}f^{\alpha}_{\sigma}{\bf G}^{r}_{\sigma}(E){\bf \Gamma}^{\alpha}_{\sigma}{\bf G}^{a}_{\sigma}(E)$ where $f^{\alpha}_{\sigma}(E)$ are Fermi distribution functions for electrons with the spin $\sigma$  on the electrodes whose temperatures and chemical potentials are $T^{\alpha}$ and $\mu_{\alpha}$, respectively, and matrices ${\bf \Gamma}^{\alpha}_{\sigma}$ describe the coupling of the triple dot to the electrodes. We compute the occupation numbers on the dots by self-consistently solving Eqs (\ref{9}).

 Within the wide band approximation ${\bf \Gamma}^{\alpha}_{\sigma}$ do not depend on the tunnel energy $E$. Also, in a system with nonmagnetic electrodes, these matrices, as well as Green's functions, are independent of the electron spin provided that the transport remains spin conserving. Assuming that dot 1 is symmetrically coupled to the electrodes, matrices ${\bf \Gamma}^{\alpha}$ for the transport junction schematically shown in Fig.1 have the form:
\begin{align}
{\bf \Gamma}^{L}=\left(\ba{ccc}{\Gamma_{1}}&{0}&{0}
\\
0&0&0
\\
0&0&0
\\
\ea\right) ;\qquad    {\bf \Gamma}^{R}=\left(\ba{ccc}{\Gamma_1}&{0}&{0}
\\
{0}&{0}&{0}
\\
0&0&{\Gamma_{3}}
\\
\ea\right) .            \label{10}
\end{align}
where the parameters $\Gamma_1^{L}=\Gamma_1^{R}\equiv\Gamma_1$ and $\Gamma_3^{R}\equiv\Gamma_3$ describe the couplings of dot 1 to both electrodes and the coupling of dot 3 to the right electrode, respectively. 
Using Eqs.(\ref{5})-(\ref{9}) we find the electron transmission function $\tau(E)=Tr\{{\bf G}^{a}(E){\bf \Gamma}^{R}{\bf G}^{r}(E){\bf\Gamma}^{L}\}$ which is reduced to $\tau(E)=\Gamma_1^{2}\big|G^{r}_{11}\big|^{2}$ in the case of the considered transport junction.

\section {III. Charge currents}.
 
Thermoelectric properties of a transport junction could be analyzed by studying charge and heat currents generated by a bias voltage and/or a temperature gradient applied across the system \cite{37,40,51,52,53,54,55,56,57}. Further we use the standard Landauer expression for the charge current $I$ \cite{4,51}:
\be
I=\frac{e}{h}\int dE\tau(E)\left(f^{L}(E)-f^{R}(E)\right).        \label{11}
\ee
Similarly, the heat current through the system may be presented in the form \cite{15,51}:
\be
J=\frac{1}{h}\int dE\tau(E) (E-\Delta\mu)\left(f^{L}(E)-f^{R}(E)\right).        \label{12}
\ee

\begin{figure}[t] 
\begin{center}
\includegraphics[width=8cm,height=6cm]{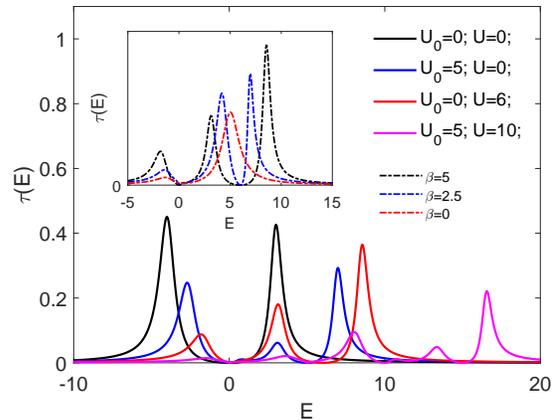}
\caption{Electron transmission through the triple dot as a function of the tunnel energy. Curves are plotted assuming $\Gamma=\Gamma_1=\Gamma_3=2$ meV, $E_{1}=-2$ meV, $E_{3}=-3$ meV, $E_{2}=0$, $\mu_L=\mu_R=0$, $\beta=10$ meV (main body);$U_0=10$ meV, $U=0$ (inset). $\Gamma$ is accepted as a unit of energy.
}
 \label{rateI}
\end{center}\end{figure}
 We first consider the electron transmission function which, besides other factors, depends on the positions of the energy levels on the dots and on inter-dot coupling and Coulomb interactions, as shown in Fig.2. One observes that in the absence of Coulomb interactions between the T-shaped portion and  dot 3 ($U=0$) the latter becomes irrelevant. Then inter-dot tunneling causes dips associated with Fano effect at $E=E_2=0$ and $E=E_2+U_0$, as displayed in the inset. These results agree with those earlier reported for T-shaped double dots (see e.g Ref.\cite {31}). If the tunneling ceases to exist quantum interference does not occur, and the transmission shows peaks at $E=E_1$ and $E=E_1+U_0$. When the charging energy $U\neq 0$ the transmission line shapes become more complicated. Besides Fano dips at $E=E_2$ and $E=E_2+U_0$ extra dips emerge at $E=E_2+U$ and $E=E_2+U_0+U$ which also originate from Fano effect. In general, we  see  that Coulomb repulsion  between electrons on  the dots noticeably reduces the transmission peaks thus opposing electron transport through the triple dot. 

Characteristics of electron transmission spectrum are reflected in the behavior of charge and heat currents which is discussed below. 
We start from a brief analysis of the charge current solely driven by a bias voltage which is supposed to be symmetrically distributed across the system ($\mu_{L,R}=\ds\pm\frac{1}{2}|e|V$). To recognize manifestations of Fano effect in $I-V$ characteristics  we first omit from consideration electron-electron interactions assuming that $U_0=U=0$ and that $E_1=E_2=0$, as well. Resulting $I-V$ curves are displayed in the inset to Fig.3. It is shown that when dot 1 and dot 2 are coupled and Fano antiresonance at $E_2=0$ emerges, a plateau centered at $V=0$ appears on the current-voltage curve. This feature vanishes when the dots are decoupled and the sole transport channel is associated with dot 1. Including into consideration Coulomb interactions we disclose similar plateaus centered at $\ds\frac{1}{2}|e|V=E_2+U_0$ and/or $\ds\frac{1}{2}|e|V=E_2+U$ which are caused by Fano effect. Instead of plateaus, regions of negative differential conductance $dI/dV$ may appear as a result of combined action of quantum interference and inter-dot Coulomb interactions.
\begin{figure}[t] 
\begin{center}
\includegraphics[width=7cm,height=5.5cm]{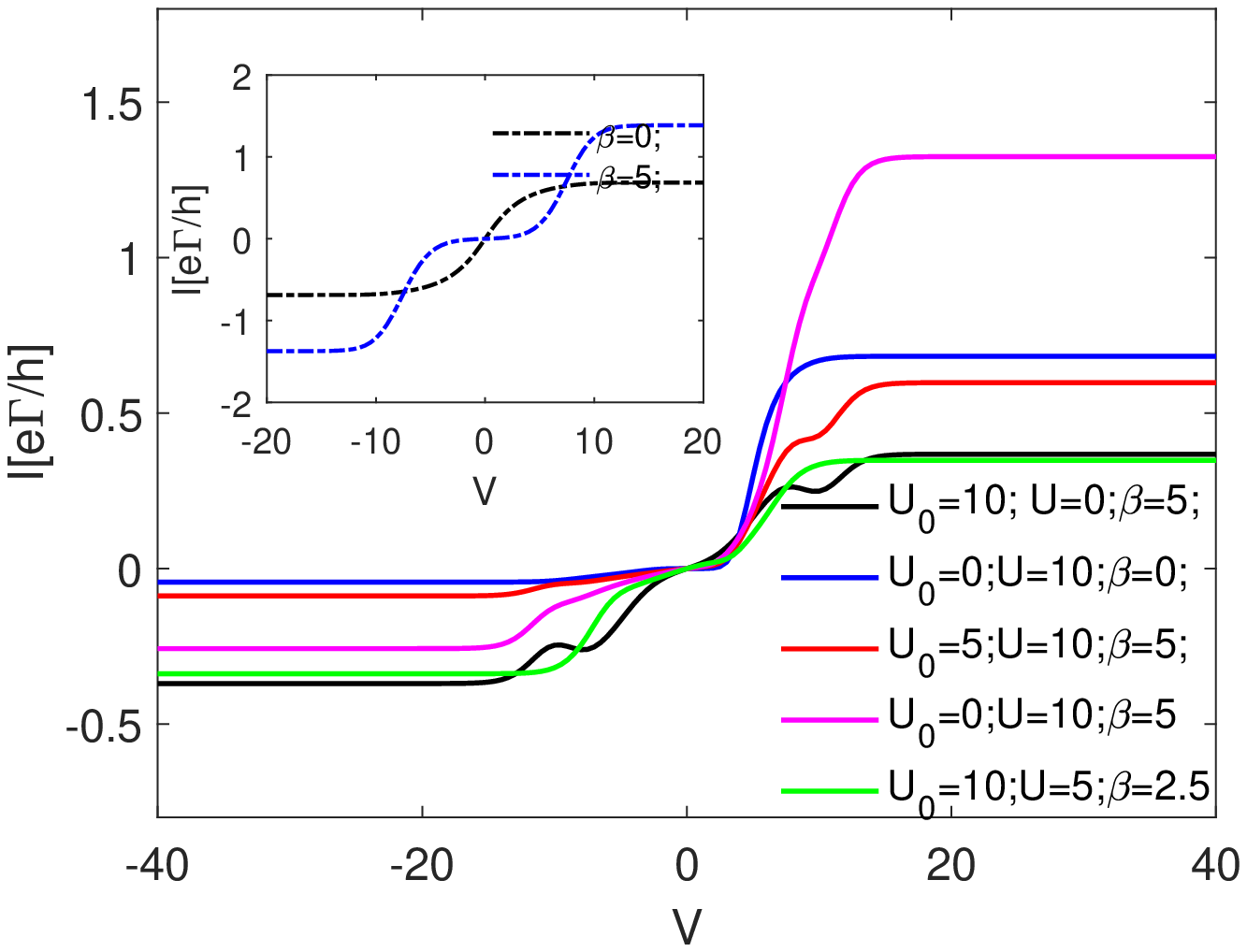} 
\includegraphics[width=7cm,height=5.5cm]{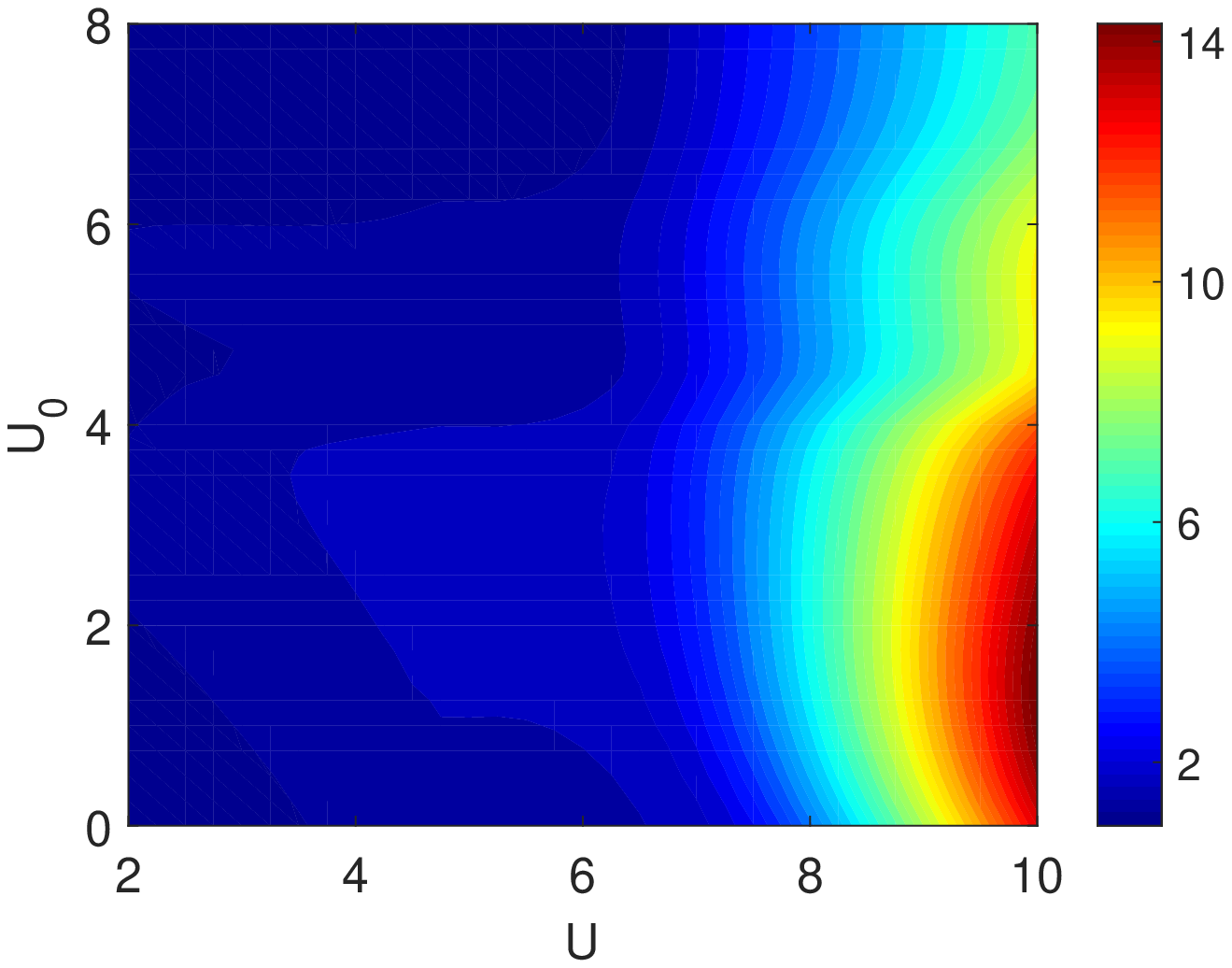}
\caption{Charge current $I$ as a funcion of the bias voltage $V$ (top) and rectification ratio $R_{I}$ as a function of charging energies $U$ and $U_0$ (bottom)  plotted assuming that $kT=1$ meV , $\Gamma=\Gamma_1=\Gamma_3=2$ meV, $E_{2}=0$ meV, $E_{3}=-3$ meV. $\Gamma$ is chosen as a unit of energy and $\Gamma/|e|$ as a unit of voltage. Top panel: $E_1=-2$ meV (main body) and $E_1=U=U_0=0$ (inset). Bottom panel:$E_1=-2$ meV, $\beta=5$ meV, $V=40$ mV. 
}
 \label{rateI}
\end{center}\end{figure}

 Coulomb repulsion between electrons on dot 3 and those on dot 1 and dot 2 makes $I-V$ characteristics asymmetric indicating rectifying properties of the considered triple dot. The rectification is especially well pronounced when the side dot is detached from dot 1 ($\beta=0$) and remains empty thus suppressing the alternative pathway for traveling electrons and quantum interference effects. In this case the triple dot is reduced to a double dot considered in the earlier work \cite{23}. In general, the rectifying properties of the system are more distinct when Coulomb repulsions between dot 3 and dots included in the T-shaped portion of the triple dot considerably exceed the inter-dot coupling ($U\gg\beta$) weakening the effect of quantum interference. Also, the current rectification is more distinct when $U_0$ is significantly smaller than $U$. At strong Coulomb repulsion between dots 1 and 2 $(U_0>\beta,U)$ $I-V$ curves remain nearly symmetrical even at $U\neq 0$. This is confirmed by the results plotted in the bottom panel of Fig.3 where we show the rectification ratio $R_{I}=\big|I_{+}/I_{-}\big|$, ($I_{+}$ and $I_{-}$ being currents at the forward and reversed bias voltage of the same magnitude) as a function of charging energies $U$ and $U_0$.  

When the electrodes temperatures differ, charge current may be jointly driven by the  bias voltage $V$ and temperature gradient $\Delta T$. We could separate out a thermally induced current (thermocurrent) $I_{th}$ defined as the difference $I(V,T,\Delta T)-I(V,T,\Delta T=0)$ ($T=\ds\frac{1}{2}(T_L+T_R)$). As follows from its definition, thermally induced portion of the charge current is rather pronounced at weak bias voltage when thermal and electrical driving forces are comparable, and it becomes negligible at stronger bias when the electrical driving exceeds the thermal driving. Further we compute $I_{th}$ assuming a symmetrical distribution of the temperature difference: $T_{L,R}=T\pm\ds\frac{1}{2}\Delta T$. 

We could detect signatures of Fano antiresonances studying dependencies of $I_{th}$ on the temperature gradient and on the bias voltage which are presented in Fig.4. Assuming that inter-dot Coulomb interactions are negligible, electron transport  strongly depends on the coupling between dots 1 and 2. Comparing  curves shown in the inset to the upper panel one sees that the inter-dot coupling results in the $I_{th}$ weakening and in the decrease of the corresponding $I_{th}-\Delta T$ curve slope at small $\Delta T$. The latter feature originates from Fano effect.

 The thermocurrent behavior becomes more complex when it is simultaneously driven by thermal and electric forces. If the dots 1 and 2 are decoupled and the sole transport channel is associated with dot 1 the corresponding $I_{th}-V$ curve plotted at a fixed negative $\Delta T$ ($T_{L}<T_{R})$ and displayed in the inset to the lower panel shows two peaks. The double-humped line shape is generated by a combined effect of the bias voltage and the thermal gradient which pushes charge carriers of both kinds (electrons and holes) from the hot (right) electrode to the cool (left) one. One peak appears when $\ds\frac{1}{2}|e|V$ is close to $E_1$. Here, the charge current is mostly maintained by electrons and controlled by the thermal gradient. Another peak emerges near $\ds\frac{1}{2}|e|V\approx kT_{R}$ where $I_{th}$ is mostly generated by holes driven by the bias voltage. In between the peaks both electrons and holes contribute to the charge flow and their contributions counterbalance each other in part or completely.  When electrons could tunnel between the dots and two peaks separated by Fano antiresonance emerge in the the electron transmission, $I_{th}-V$ line shape becomes different. New features appear at $\ds\frac{1}{2}|e|V\approx\tilde E_{1,2}$ indicating several changes in the $I_{th}$ direction. Note that the thermocurrent changes its direction at $V=0$ and remains weak over a certain interval around this point. This behavior is also a manifestation of Fano effect as well as the reduced slope of $I_{th}-\Delta T$ curve at small $\Delta T$ and the appearance of plateau around $V=0$ in the $I-V$ curve discussed above.
\begin{figure}[t] 
\begin{center}
\includegraphics[width=7cm,height=5.5cm]{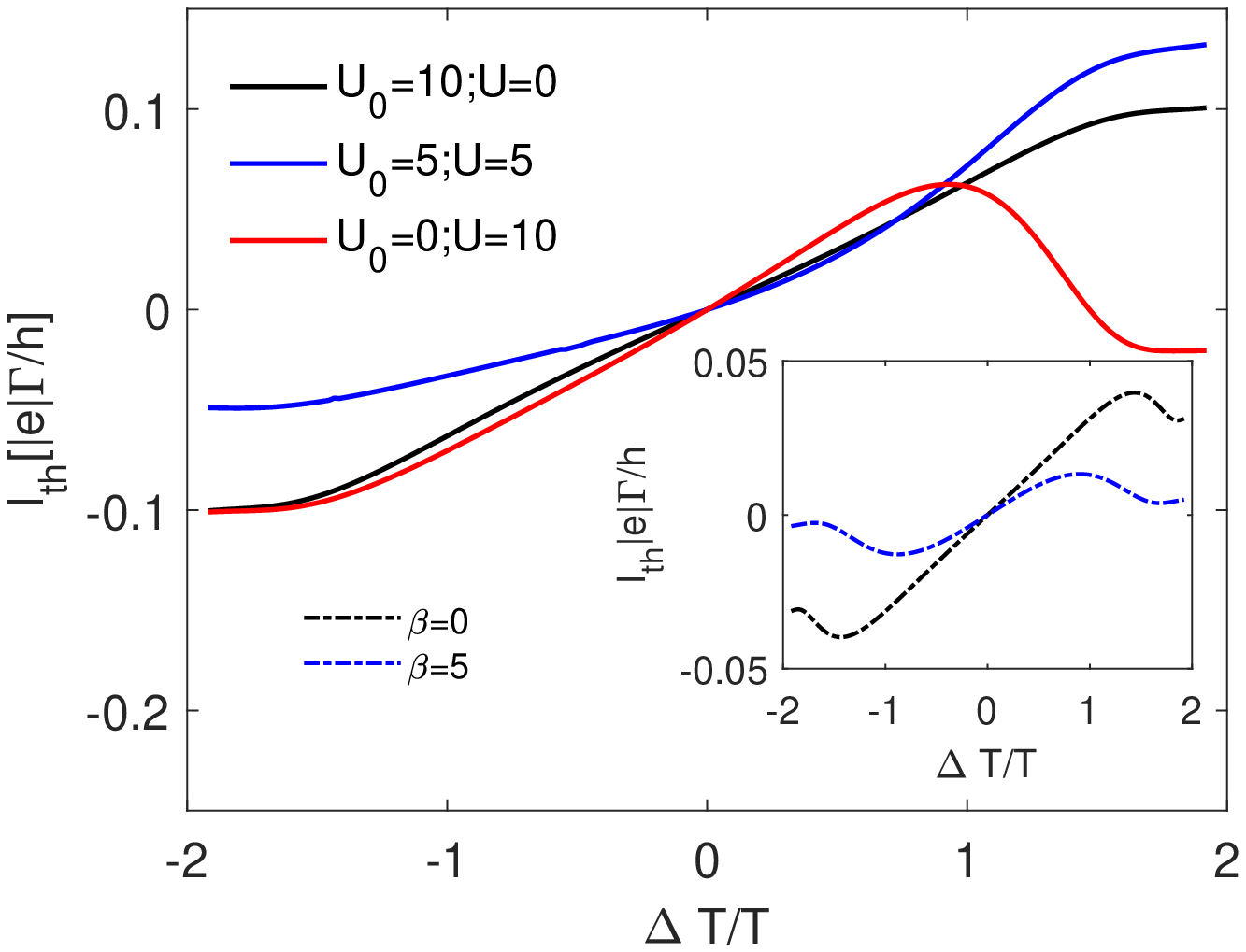}
\includegraphics[width=7cm,height=5.5cm]{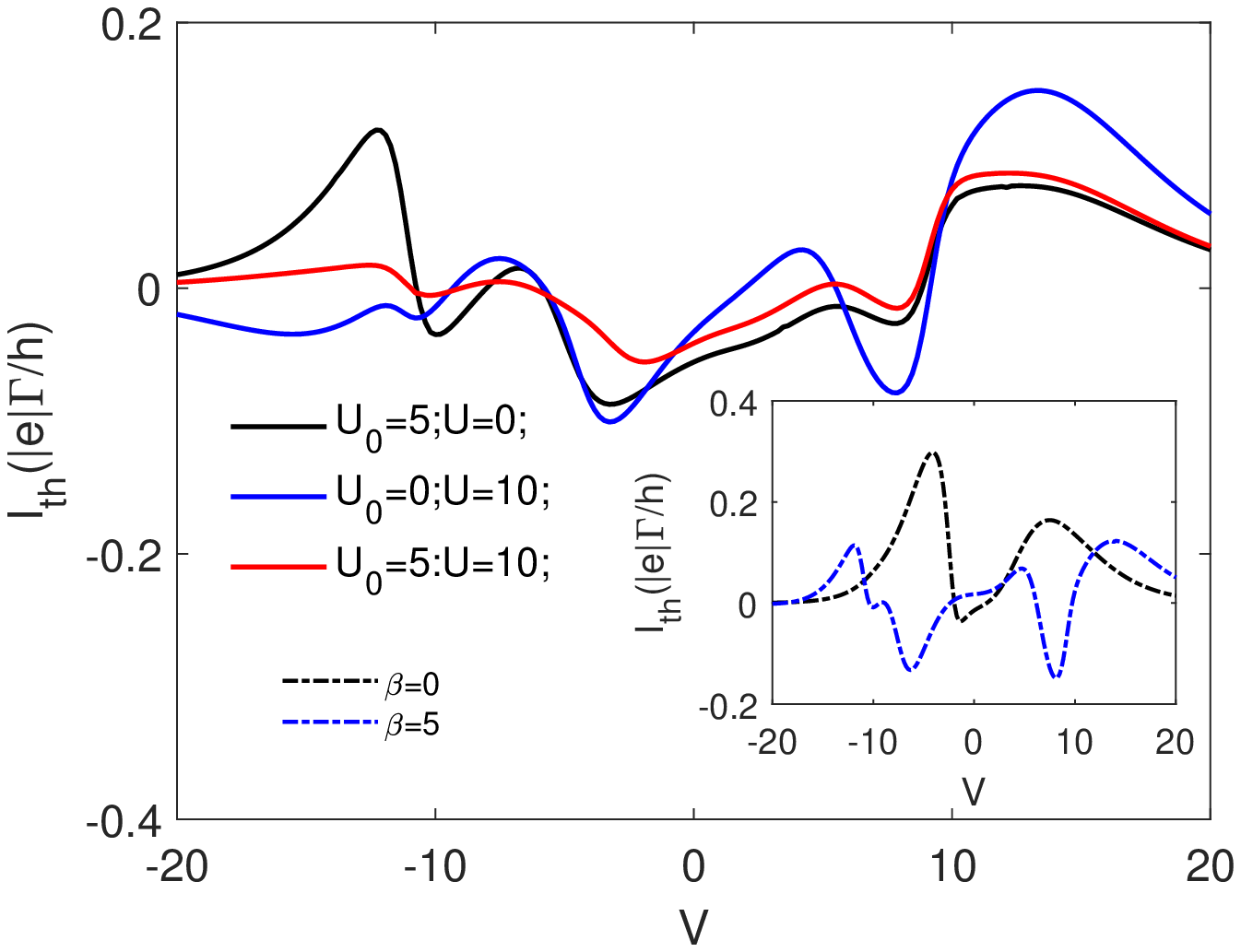} 
\caption{Thermally induced current $I_{th}$  versus temperature difference in the absence of a bias voltage (top) and $I_{th}$ versus bias voltage $V$ at a fixed temperature difference $\ds\frac{\Delta T}{T}=-1.6$ (bottom). Curves are plotted assuming that $kT=5$ meV (top), $kT=2$ meV (bottom), $\Gamma=\Gamma_1=\Gamma_3=2$ meV, $E_{1}=-2$ meV, $E_{2}=0$, $E_{3}=6$ meV, $\beta=10$ meV (main bodies), $U_0=U=0$ (insets). $\Gamma$ and $\Gamma/|e|$ are used as units of energy and voltage, respectively.
}
 \label{rateI}
\end{center}\end{figure}

 Inter-dot Coulomb interactions cause changes in $I_{th}$ behavior. These changes are especially notable when interactions between electrons on dot 3 and those on the remaining dots occur, for they are responsible for asymmetry of $I_{th}$ versus $\Delta T$ and $I_{th}$ versus $V$ curves. At negative temperature gradient and/or bias voltage  all three dots are occupied, and Coulomb repulsion between dot 3 and dots 1 and 2 partly suppresses the thermocurrent, whereas at positive and sufficiently strong $V$ and/or sufficiently pronounced $\Delta T$ the occupation of dot 3 drops and the effect of Coulomb interactions weakens, as shown in Fig.4.   

\begin{figure}[t] 
\begin{center}
\includegraphics[width=7cm,height=5.5cm]{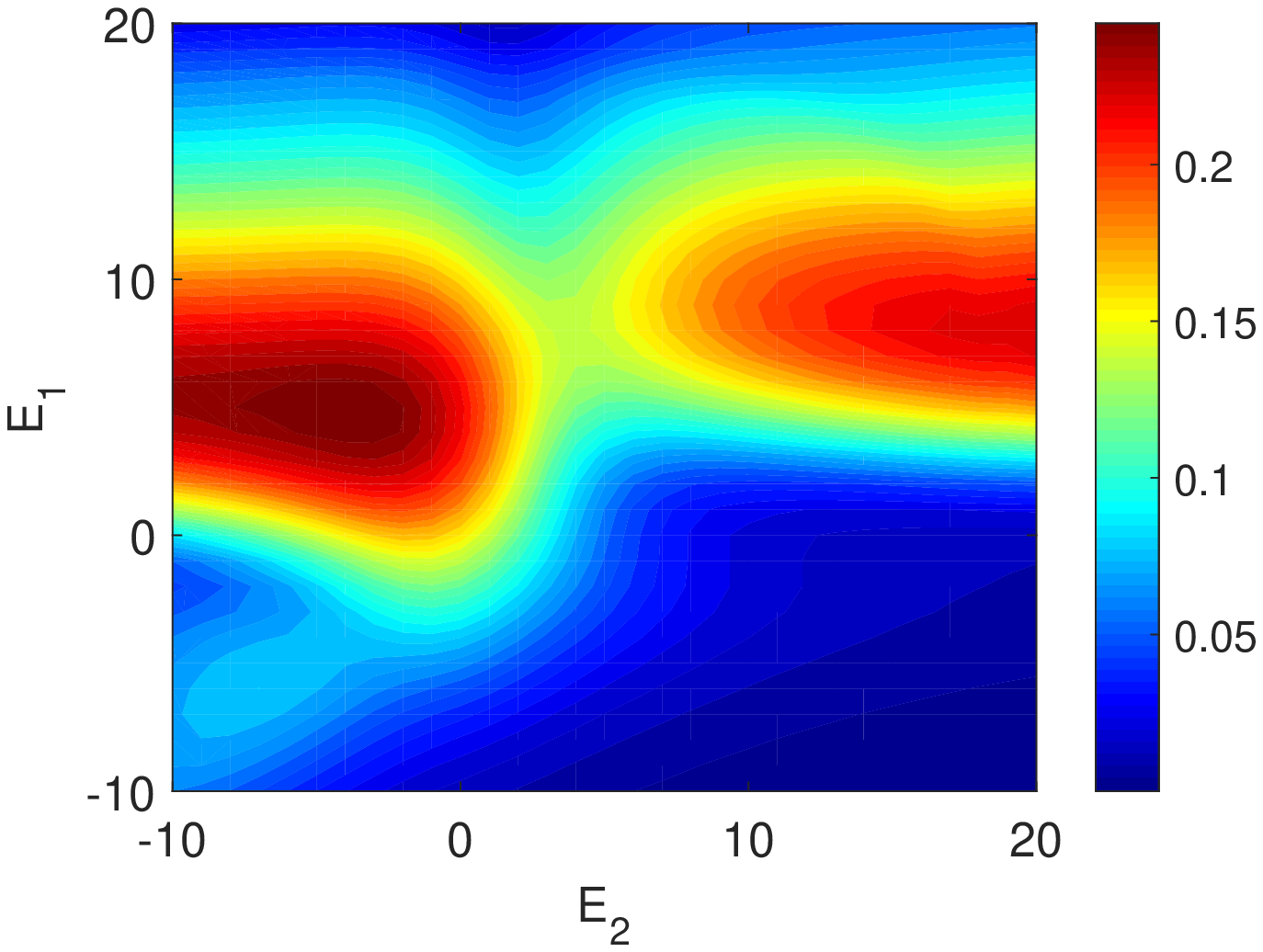} 
\includegraphics[width=7cm,height=5.5cm]{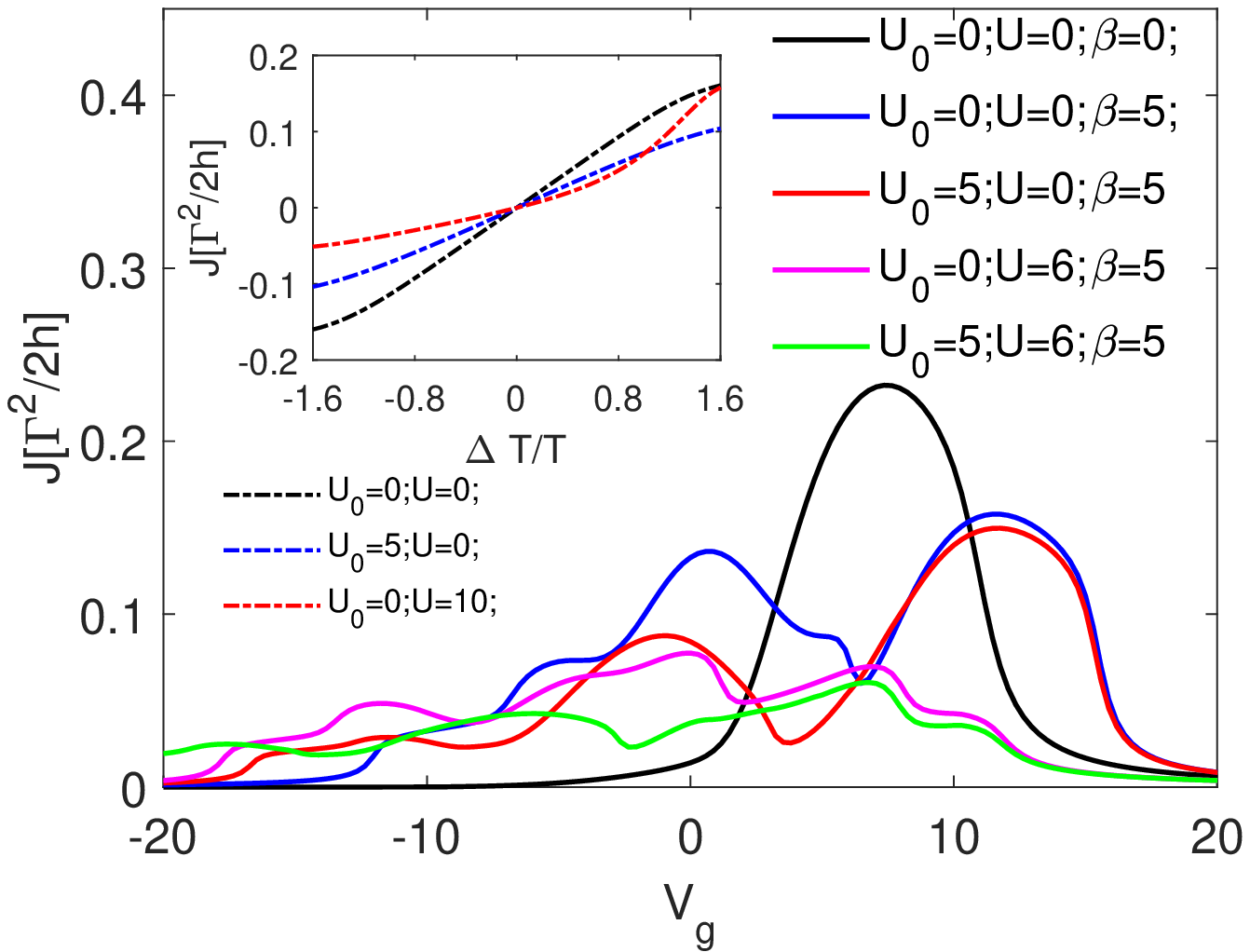}
\caption{Heat current $J$ as a function of the on-site energies $E_1$ and $E_2$ (top) and of the gate voltage $V_g$ and the temperature gradient $\Delta T$ (bottom)  plotted  at $kT=5$ meV, $\Gamma=\Gamma_1=\Gamma_3=2$ meV. In the top panel $\ds\frac{\Delta T}{T}=1.2$, $E_{3}=6$ meV, $\beta=10$ meV, $U_0=U=0$. In the bottom panel $E_1=-2$ meV, $E_2=0$, $E_3=-3$ meV, $\ds\frac{\Delta T}{T}=1.2$ (inset). $\Gamma$ is used as the unit of energy.
}
 \label{rateI}
\end{center}\end{figure}

\section{IV. Heat currents.}

Now, we turn to studies of heat currents flowing through the system biased by the applied temperature gradient $\Delta T$. Again, we assume a symmetric distribution of the temperature between the electrodes, so $T$ remains constant. We compute the heat current using the general expression given by Eq.(\ref{12}).  Adopting this expression we omit from consideration phonon mechanisms of energy transfer and concentrate on the electron contribution to the heat current.
In the considered case, the difference in the chemical potentials of electrodes $\Delta\mu=\mu_L-\mu_R$  occurs due to the thermovoltage $\Delta V_{th}$ which  originates from Seebeck. As before, we consider the electron transport within the Coulomb blockade regime. 

To better elucidate how  quantum interference may affect the heat current we first consider a case simplified by omitting Coulomb interactions. In this case, the dip in the electron transmission resulting from Fano effect occurs at $E_2=\mu$ ($\mu$ being the chemical potential of electrodes in the absence of bias voltage). However, charge carriers whose energy equals $\mu$ do not contribute to the heat flow from the hot electrode to the cool one provided that $\mu=0$. Assuming that $T_{L}>T_{R}$, the heat current is mostly generated by electrons with energies slightly shifted towards $ kT_{L}$, and we may expect a Fano effect related minimum in the heat current to appear at $E_2\sim kT_{L}$, as shown in Fig.5 (see bottom panel). The minimum remains distinguishable when $E_1$ is sufficiently close to $E_2$.

\begin{figure}[t] 
\begin{center}
\includegraphics[width=5cm,height=5cm]{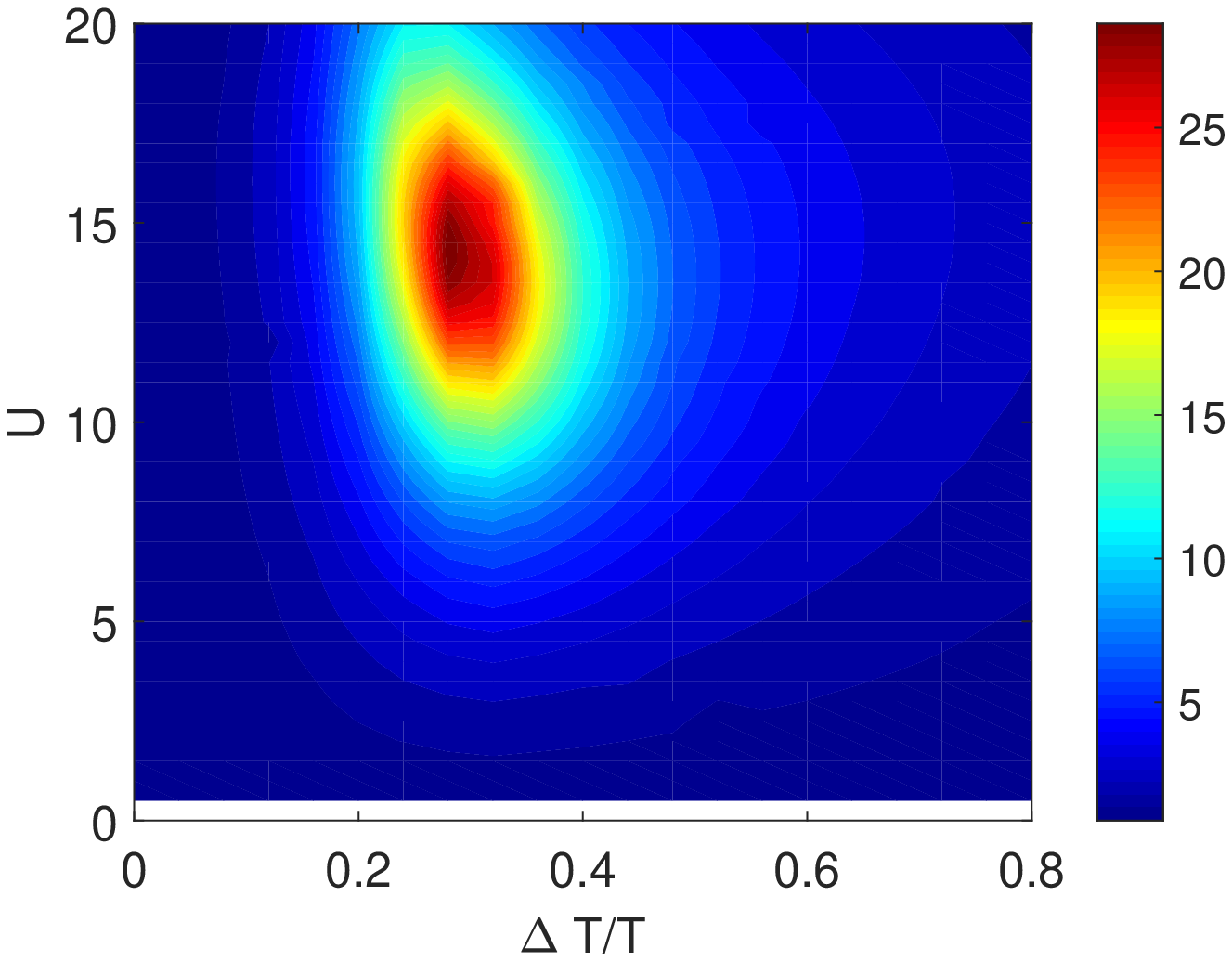} 
\includegraphics[width=5cm,height=5cm]{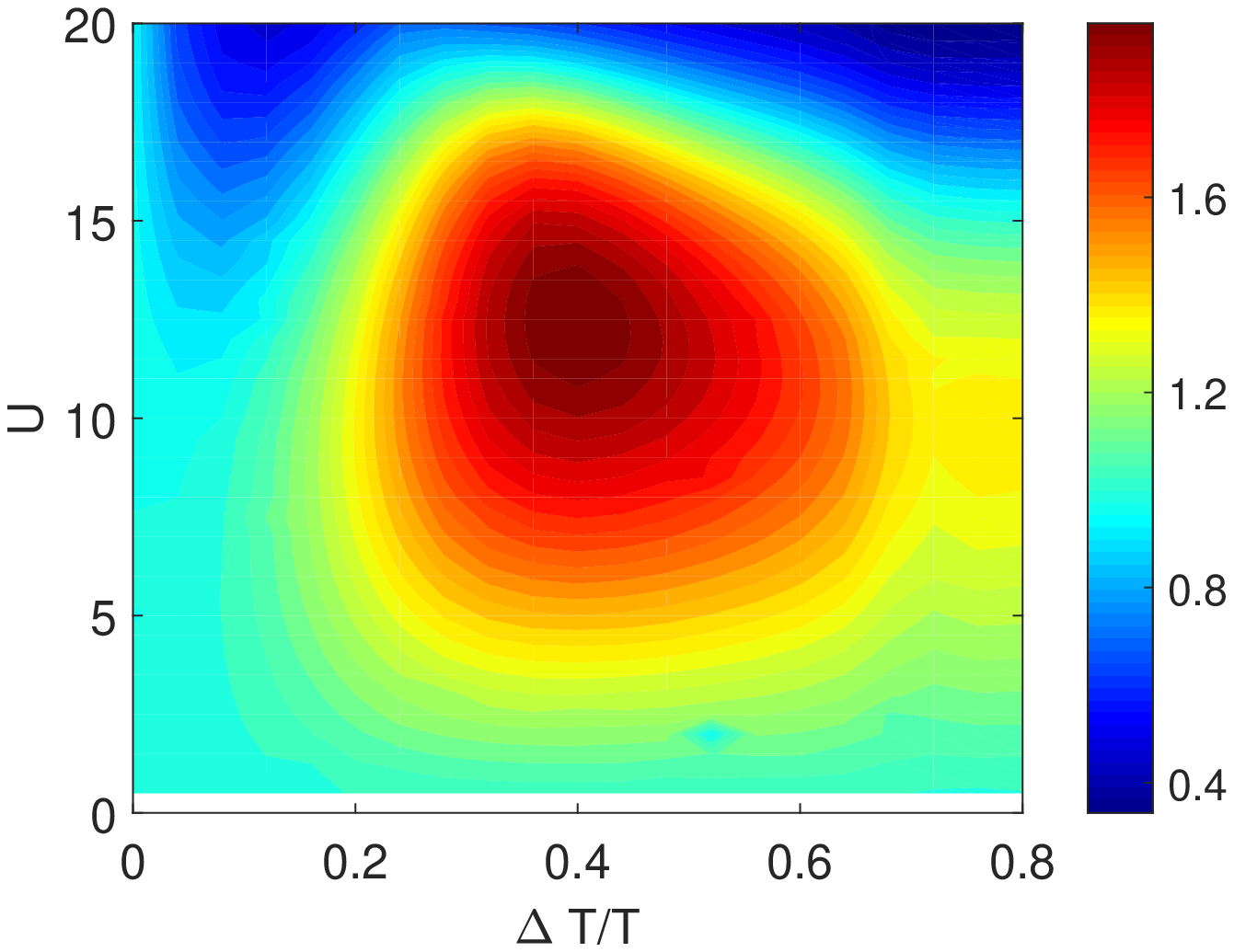}
\includegraphics[width=5cm,height=5cm]{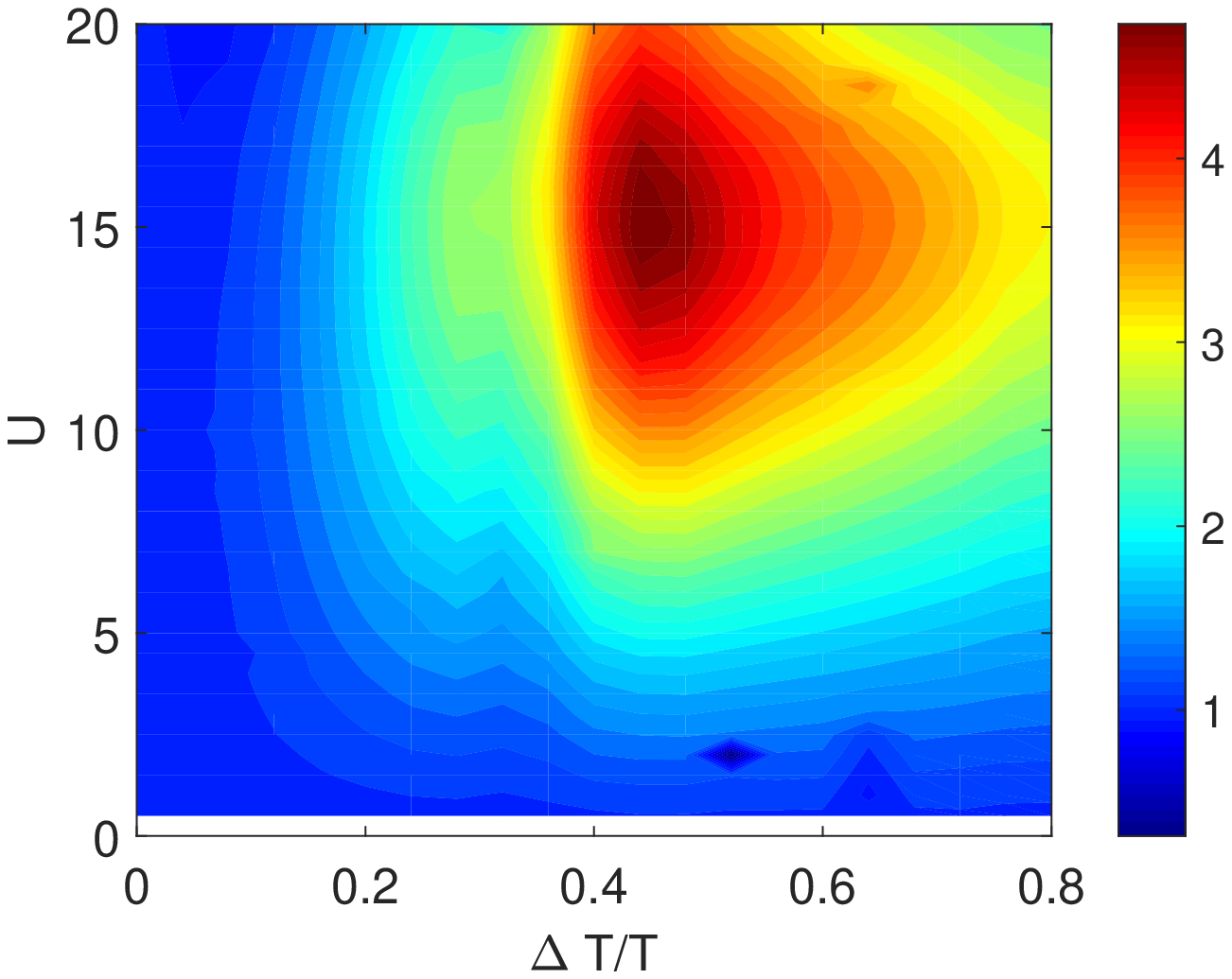}
\caption{Heat current rectification ratio $R_{J}$ as a function of the temperature gradient $\Delta T$ and of the charging energy $U$ plotted at $kT=5$ meV, $E_1=-2$ meV, $E_2=0$, $E_3=6$ meV, $\Gamma=\Gamma_1=\Gamma_3=2$ meV. Top and middle panels are plotted assuming that $U_0=0$ and $\beta=5$ meV (top panel) and $\beta=10$ meV (middle panel). Bottom panel is plotted assuming that $\beta=U_0=10$ meV. $\Gamma$ is used as the unit of energy.
}
 \label{rateI}
\end{center}\end{figure}

 Signatures of Fano dips in the transmission may be discovered by analyzing the dependencies of the heat current on the gate voltage also displayed in Fig.5. Provided that $U_0=U=0$ and the dots 1 and 2 are decoupled, the heat current shows a single maximum at $V_g\sim E_1+kT_{L}$ which occurs due to the electron transport via dot 1. Coupling between the dots results in splitting of this peak in two and emerging of a minimum indicating Fano antiresonance at $V_g\sim kT_{L}$. Coulomb repulsion between dots 1 and 2 ($U_0\neq 0$) results in occurring of two dips arising from the Fano effect, and the repulsion between electrons on these dots and those on dot 3 ($U_0\neq 0$, $U\neq 0$) partially suppresses  the heat current and brings additional features into $J$ versus $V_g$ line shapes.

The dependence of $J$ on the thermal gradient is presented in the inset to the bottom panel of Fig.5. One observes that Coulomb interactions noticeably suppress the heat transfer through the considered system and may bring asymmetry into $J$ versus $\Delta T$ curves which indicates heat current rectification. It was suggested in some earlier works \cite{22,23,29,58,59,60} that inter-dot Coulomb interactions may cause  rectification of a heat current in a double QD. One may expect that the same could happen in the considered triple dot provided that Coulomb repulsion between dot 3 and dots 1 and 2 is sufficiently strong. In Fig.6 we show the rectification ratio for the heat current $R_{J}=\big|J_{+}/J_{-}\big|$ ($J_{\pm}$ being heat currents corresponding to the forward $\Delta T>0$ and reversed $\Delta T<0$ bias created by the thermal gradient). In this figure, left and middle panels are plotted assuming that $U_0=0$ for two different values of the coupling parameter $\beta$. Comparing them one sees that $R_{J}$ may reach values of the order of 10 provided that inter-dot tunneling in the T-shaped portion is weak. However, a stronger tunneling significantly worsens rectifying properties of the triple dot. The effect of inter-dot tunneling and associated quantum interference may be partially neutralized by Coulomb repulsion between dots 1 and 2 for it opposes the inter-dot tunneling. This is illustrated in the bottom panel of Fig.6 which is plotted accepting the same value of $\beta$ as that in the middle panel but using $U_0\neq 0$.

\section {VI. Conclusion.}

Single and multiple QDs are being manufactured and studied both theoretically and experimentally. In the present work we theoretically analyzed thermoelectric transport through a system consisting of a $T$-shaped double dot completed with a third dot in a parallel configuration and placed in between nonmagnetic electrodes. We studied electron transport through the chosen triple QD within the Coulomb blockade regime. Effects of electron-phonon interactions on the electron transport were omitted basing on the assumption that phonon contributions to both electrical and thermal conductance of the system were small and coherent electron transport predominated. Transport characteristics were computed using the Green's functions formalism within the limit of strong intra-dot Coulomb interactions where intra-dot charging energies greatly exceeded inter-dot ones.

We focused on the effect of quantum interference between the transport channels provided by the $T$-shaped portion of the triple QD on the thermoelectric transport beyond the linear response regime. It was shown that Fano effect may affect charge and heat currents flowing through the considered transport junction. The effect of quantum interference strongly depends on electron energies on the dots included in the $T$-shaped portion of the triple dot and of inter-dot coupling strength which determines positions of Fano antiresonances on the energy scale. 

Electron-electron interactions between the dots may result in rectification of charge currents driven by a bias voltage as well as heat currents driven by the thermal gradient. We showed that in both cases the corresponding rectification ratio is controlled by the interplay between inter-dot electron-electron interactions and quantum interference effects associated with electron tunneling in the T-shaped portion of the considered triple QD. Specifically, inter-dot tunneling may significantly reduce the rectification ratio characterizing both charge and heat currents. 

Adopted model of a triple dot may be employed to better understand certain aspects of electron transport in metal-molecule transport junctions where several pathways for traveling electrons could coexist and interference effects may affect transport characteristics together with electron-electron interactions. Therefore we believe that the presented results provide a step towards further understanding and modeling of thermoelectric and heat transport through quantum dots and molecules.

\section{Data availability statement}

Data sharing is not applicable as no new data are created in this study.

\section {Acknowledgements.} This work was supported by NSF-DMR-PREM 2122102. The author thanks G. M. Zimbovsky for help with the manuscript preparation


\begin{thebibliography}{99}



\bibitem{1} Agrait, N.; Yeati, A. L.; Van Ruitenbeck, J. M., {\it Phys. Rep.} {\bf 2003}, 377, 81-273.

\bibitem{2} Giazotto, F.; Heikkila, T. T.; Luukanen, A.;  Savin, A. M. and Pecola, J. P., {\it Rev. Mod. Phys.}, {\bf 2006}, 78, 217.

\bibitem{3} Dubi, Y. and Di Ventra, M., {\it Rev. Mod. Phys.}, {\bf 2011}, 83, 131.

\bibitem{4} Galperin, M.; Ratner, M. A. and Nitzan, A., {\it J. Phys.:Condens. Matter}, {\bf 2007}, 19, 103201.

\bibitem{5} Zimbovskaya, N. A. and Pederson, M. R., {\it Phys. Rep.}, {\bf 2011}, 509, 1-89.

\bibitem{6} Jezouin, S.;  Parmentier, F. D.; Anthore, A.; Gennser, U.; Cavanna, A.: Jiu, Y. and Pierre, F., {\it Science}, {\bf 2013}, 342, 601.

\bibitem{7} Riha, C.; Miechowski, P.; Buchholz, S. S.; Chiatti, O.; Wieck, A. D.; Reuter, D. and Fisher, S. S., {\it Appl. Phys. Lett.}, {\bf 2015}, 106, 083102.

\bibitem{8} Cui, L.; Jeong, W.; Hur, S.; Matt, M.; Klockner, J. C.; Pauly, F.; Nielaba, P.; Cuevas, J. C.; Meyhofer, P. and Reddy, P., {\it Science}, {\bf 2017},355, 1192.

\bibitem{9} Sothmann, B.; Sánchez, R. and Jordan, A. N.,{\it Nanotechnology}, {\bf 2015}, 26, 032001.

\bibitem{10} Saleghi, H.; Sangtarash, S. and Lambert, C. J., {\it Nano Lett.}, {\bf 2015}, 15, 7467.

\bibitem{11} Noori, M.; Saleghi, H. and Lambert, C. J., {\it Nanoscale}, {\bf 2017}, 9, 5299.

\bibitem{12} Perroni, C. A.; Ninno, D. and Cataudella, V., {\it J. Phys.: Condens. Matter}, {\bf 2016}, 28, 373001.

\bibitem{13} Lukkebo, J.; Romano, G. et al, {\it J. Chem. Phys.}, {\bf 2016}, 144, 114310.

\bibitem{14} Kulguir, M. and Segal, D., Phys. Rev. E {\bf 2018}, 98, 012114.

\bibitem{15} Kuo,  D. M.-T. and Chang, Y.-C., {\it Phys. Rev. B}, {\bf 2010}, 81, 205301.

\bibitem{16} Marcos-Vicioso, A.; López-Jurádo, C.; Rúiz-García, M. and Sánchez, R.,{\it  Phys. Rev. B}, {\bf 2018}, 98, 035414.

\bibitem{17} Segal, D. and Nitzan, A., {\it Phys. Rev. Lett.}, {\bf 2004}, 94, 034301.

\bibitem{18} Wu, L.-A.; Yu, C.-X. and Segal, D., {\it Phys. Rev. E}, {\bf 2009}, 80, 041103 (2009).

\bibitem{19} Ojanen, T., {\it Phys. Rev. B}, {\bf 2009}, 80, 180301.

\bibitem{20} Craven, G. T.; He, D. and Nitzan, A., {\it Phys. Rev. Lett.}, {\bf 2018}, 121, 247704 (2018). 

\bibitem{21} Aligia, A. A.; Péres Daroca, D.; Arrachea, L.; Roura-Bas, P., {\it Phys. Rev. B}, {\bf 2020}, 101,075417.

\bibitem{22} Zimbovskaya, N. A., {\it J. Chem. Phys.}, {\bf 2020}, 153, 124712.

\bibitem{23} Zimbovskaya, N. A., {\it J. Phys.: Condens. Matter}, {\bf 2020},32, 325302.

\bibitem{24} Di Vincenzo, D. P., {\it Science}, {\bf 2005}, 309, 2174.

\bibitem{25} Jiang, Z.-T.; Sun, Q.-F., {\it J. Phys.: Condens. Matter}, {\bf 2007}, 19, 156213.

\bibitem{26} Yang, X.-F. and Liu, Y.-S., {\it Nanoscale Res. Lett.}, {\bf 2010}, 5, 1228-1235.

\bibitem{27} Ladrón de Guevara, M. L.; Lara, G. A. and Orellana, P. A., {\it Physica E}, {\bf 2010}, 42, 1637-1642.

\bibitem{28} Lu, H.; Lü, R. and Zhu, B.-f., {\it Phys. Rev. B}, {\bf 2005}, 71, 235320.

\bibitem{29} Trocha, P. and Barnás, J., {\it  Phys. Rev. B}, {\bf 2007}, 76, 165432.

\bibitem{30} Ladrón de Guevara, M. L.; Claro, F. and Orellana, P. A., {\it Phys. Rev. B}, {\bf 2003}, 67, 195335.

 \bibitem{31} Monteros, A. L.; Uppal, G. S.; McMillan, M.; Crisan, M.; Tifrea, I., {\it Euro. Phys. J. B}, {\bf 2014},87, 302.

\bibitem{32} Orellana, P. A.; Ladrón de Guevara, M. L. and Claro, F., {\it  Phys. Rev. B}, {\bf 2004}, 70, 233315.

\bibitem{33} Trocha, P. and Barnás, J., {\it  Phys. Rev. B}, {\bf 2012}, 85, 085408.

\bibitem{34} Swírkowicz, R.; Wíerzbicki, M. and Barnás, J., {\it  Phys. Rev. B}, {\bf 2009}, 80, 195409.

\bibitem{35} Dubi, Y. and Di Ventra, M., {\it Phys. Rev. B.}, {\bf 2009}, 79, 081302.

\bibitem{36} Trocha, P. and Barnás, J., {\it  Phys. Rev. B}, {\bf 2017}, 95, 165439.

\bibitem{37} Estrada-Saldana, J. C; Verkis, A.; Steffensen, G.; Zitko, R.; Krogstrup, P.; Paaske, J.; Grove-Rasmussen, K. and Nygard, J., {\it Phys. Rev. Lett.}, {\bf 2018}, 121, 257701.

\bibitem{38} Baránski, J.; Zíenkewics, T.; Baránska, M. and Karcia, K. P., {\it Sgi. Rep.},{\bf 2020},10, 2881.

\bibitem{39} Xu, W.-P.; Zhang, Y.-Y.; Wang, Q.; Li, Z.-J. and Nie, Y.-H., {\it Phys. Lett. A}, {\bf 2016}, 380, 958.

\bibitem{40} Wíerbicki, M. and Swírcowicz, R., {\it Phys. Rev. B}, {\bf 2011}, 84, 075410.

\bibitem{41} Siérra, M. A.; Sáiz-Bretín, M.; Domínguez-Adame, F. and Sánchez, D., {\it Phys. Rev. B}, {\bf 2016}, 93, 235452.

\bibitem{42} Li, R. X.; Ni, Y.; Li, H. D.; Tian, X. L.; Yao, K. L.; Fu, H. H., {\it Physica B}, {\bf 2016}, 493, 1.

\bibitem{43} Gonzalez, A.; Pacheco, M.; Calle, A. M.; Siqueira, E. S. and Orellana, P. A., {\it Sci. Rep.}, {\bf 2021}, 11, 3941.

\bibitem{44} Calle, A. M.; Pacheco, M.; Martins, G. B.; Apel, V. M.; Lara, G. A. and Orellana, P. A., {\it J. Phys.: Condens. Matter}, {\bf 2017}, 29, 135301.

\bibitem{45} Trocha, P. and Barnás, J., {\it  Phys. Rev. B}, {\bf 2008}, 78, 075424.

\bibitem{46} Karwacki, L. and Trocha, P., {\it  Phys. Rev. B}, {\bf 2016}, 94, 085418.

\bibitem{47} Tsaousidou, M. and Tribertis, G. P., {\it J. Phys.: Condens. Matter}, {\bf 2010}, 22, 355304.

\bibitem{48} Bergfield, J. P.; Solis, M. and Stafford, C. A., {\it ACS Nano}, {\bf 2010}, 4, 5314.

\bibitem{49} Chang, Y.-C. and Kuo, D. M.-T., {\it  Phys. Rev. Lett.}, {\bf 2007}, 99, 086803.

\bibitem{50} Chang, Y.-C. and Kuo, D. M.-T., {\it  Phys. Rev. B}, {\bf 2008}, 77, 245412.

\bibitem{51} Datta, S., {\it Quantum Transport: From Atom to Transistor}, (Cambridge University Press, Cambridge, UK, 2005).

\bibitem{52} Dubi, Y. and Di Ventra, M., {\it Nano Lett.}, {\bf 2009}, 9, 97.

\bibitem{53} Leijinse, M.; Wegewijs, M. R. and Flensgerg, K.,{\it Phys. Rev. B}, {\bf 2010}, 82, 045412.

\bibitem{54} Argüello-Luengo, J.; Sánchez, D.; Lopez, R., {\it Phys. Rev. B}, {\bf 2015}, 91, 165431.

\bibitem{55} Azema, J.; Lombardo, P. and Dare, A. M., {\it Phys. Rev. B}, {\bf 2014}, 90, 205437. 

\bibitem{56} Sierra, M. A. and Sánchez, D., {\it Phys. Rev. B}, {\bf 2014}, 90, 115313.

\bibitem{57} Svensson, S. F.; Hoffmann, E. A.; Nakpathomkin, N.; Wu, P. M.; Xu, H. Q.; Nilsson, H. A.; Sánchez, D.; Kashcheyevs, V. and Linke, H., {\it New J. Phys.}, {\bf 2013}, 15, 105011.

\bibitem{58} Ruokola, T.; Ojanen, T., {\it Phys. Rev. B}, {\bf 2011}, 83, 241404(R).

\bibitem{59} Roselló, G., López, R. and Sánchez, R., {\it Phys. Rev. B}, {\bf 2017}, 95, 235404.

\bibitem{60} Sánchez, R.; Thierschmann, H. and Molencamp, L. W., {\it New J.. Phys.}, {\bf 2017}, 19, 113040.


\end{thebibliography}
\end{document}